\documentclass[11pt,,epsf]{article}
\usepackage{graphicx}
\begin{document}
\centerline{\bf\Large Can the observed enhancement in the mass
spectrum of $p\bar p$ in $J/\psi\rightarrow \gamma p\bar p$}
\vspace{0.3cm} \centerline{\bf\Large be interpreted by a possible
$p\bar p$ bound state}

\vspace{1cm}

\centerline{Xiang Liu$^1$, Xiao-Qiang Zeng$^1$, Yi-Bing
Ding$^{6,3}$, Xue-Qian Li$^{1,2,6}$, Hong Shen$^1$ and Peng-Nian
Shen$^{5,4,2,6}$}

\vspace{0.4cm}

\noindent 1. Department of Physics, Nankai University, Tianjin
300071, China\\

\noindent 2. Institute of Theoretical Physics, CAS, P.O. Box 2735,
Beijing, 100080, China\\

\noindent 3. Graduate School of The Chinese Academy of Sciences,
Beijing, 100039, China\\

\noindent 4. Institute of High Energy Physics, CAS, P.O. Box
918(4), Beijing 100039, China\\

\noindent 5. Center of Theoretical Nuclear Physics, National
Laboratory of Heavy Ion Accelerator, Lanzhou 730000, China\\

\noindent 6. China Center of Advanced Science and Technology
 (World Laboratory), P.O.Box 8730, Beijing 100080, China

\vspace{1cm}

\begin{center}
\begin{minipage}{12cm}
\noindent {\bf Abstract}

Provided the enhancement in the $p\bar{p}$ spectrum in radiative
decay $J/\psi\rightarrow\gamma p\bar{p}$ observed by the BES
collaboration is due to an existence of a $p\bar{p}$ molecular
state, we calculate its binding energy and lifetime in the linear
$\sigma$ model. We consider a possibility that the enhancement is
due to a $p\bar p$ resonance which is in either S-wave or P-wave
structure and compare our results with the data.

\end{minipage}
\end{center}

\vspace{1cm}

\section{Introduction}

\hspace{6mm}Recently, the BES collaboration has observed a
near-threshold enhancement in the $p\bar{p}$ mass spectrum in the
radiative decay $J/\psi\rightarrow \gamma p \bar{p}$ \cite{BES}. A
similar report about the enhancement in $\bar{B^{0}}\rightarrow
D^{*0} p \bar{p}$ and $B^{\pm}\rightarrow p \bar{p} K^{\pm}$
decays has been published by the Belle collaboration \cite{Belle}.

There have been various interpretations for the observed
enhancement. The enhancement can be understood if the final state
interaction between $p$ and $\bar{p}$ is properly considered, as
some authors suggested \cite{Zou}. He et al. propose a possible
mechanism that the final state of $\gamma p\bar{p}$ comes from an
intermediate state of $\gamma+G$ where G is a $0^{-+}$ or $0^{++}$
glueball \cite{He}. Meanwhile in analog to $a_{0}(980)$ and
$f_{0}(980)$ which are supposed to be molecular states of
$K\bar{K}$, it is tempted to assume that $p\bar{p}$ constitute a
bound state with quantum number $0^{-+}$ or $0^{++}$.

Under the assumption, one needs to evaluate the corresponding
binding energy and lifetime, then compare the theoretical results
with the data. In this work, we employ the linear $\sigma$ model.
Historically, there has been dispute about the linear $\sigma$
model where the $\sigma$-meson stands as a realistic scalar meson
\cite{Georgi} whereas in alternative scenarios, it is suggested
that the contribution of $\sigma$ can be attributed to two-pion
exchange \cite{Rept}. In fact, the difference between the linear
$\sigma$ model and non-linear $\sigma$ model is whether the
$0^{++}$ $\sigma$-meson is a substantial object, it corresponds to
the linear or non-linear realization of the chiral lagrangian
\cite{Wu}. Because the low-energy QCD which is the underlying
theory of hadron physics is fully non-perturbative, all the
nonperturbative QCD parameters in the theory are so far not
strictly derivable and have to be extracted by the data fitting.
Therefore, determination of these parameters is somehow
model-dependent and phenomenological. It is believed that at least
for the leading order, all models would be applicable, even though
they look somewhat different. As we employ the linear $\sigma$
model which is simpler in calculations, we take all the
coefficients by fitting data.

In our earlier work \cite{Liu}, we used the linear $\sigma$ model
to calculate the properties of deuteron, and by fitting data we
not only determine the value of $m_{\sigma}$ but also fix the
corresponding parameters of the linear $\sigma$ model. In this
work we will use the same model with the parameters obtained by
fitting the deuteron data to carry out all calculations for the
$p\bar{p}$ bound state.

The present BES data do not finally decide if the resonance is an
S-wave or P-wave bound state, but only indicate that the position
of the S-wave peak is below the threshold $2m_{p}$ whereas the
peak of the P-wave is a bit above the threshold. In our model,
since the effective potential for the S-wave is attractive except
a repulsive core near $r\rightarrow 0$, the binding energy must be
negative, so that the calculated mass of the S-wave bound state is
below the $2m_{p}$ threshold. Whereas for the P-wave due to the
angular momentum barrier which is non-zero and positive, the
binding energy becomes positive and the total mass is greater than
$2m_{p}$.

To evaluate the widths for both S-wave and P-wave, we investigate
the dissociation mechanism of the $p\bar p$ bound state. Since the
central value of the mass of the S-wave bound state is smaller
than $2m_{p}$, it dissolves into a $p\bar{p}$ pair via its width
tail where the available energy is sufficient to produce a free
$p\bar{p}$ pair. To evaluate the total width of the bound state,
we need to achieve the imaginary part of the potential which is
induced by the absorptive part of the loops in the $p\bar p$
elastic scattering amplitude (see the text for the concerned
Feynman diagrams and some details). Thus according to the
traditional method \cite{Landau}, we derive the real part of the
potential which mainly comes from the tree-level scattering
amplitude where t-channel mesons are exchanged, including
$\sigma$, $\pi$, $\rho$ and $\omega$. For the S-wave bound state
not only t-channel exchange, but also the s-channel annihilation
contribute. Namely in the s-channel, $\eta$ and $\eta'$ are the
intermediate mesons and they contribute a real part  and an
imaginary part to the effective potential, the s-channel
contributions are proportional to a delta function $\delta({\bf
r})$ in the non-relativistic approximation. Thus the eigenenergy
becomes $E_{Re}-i\frac{\Gamma}{2}$ and the time-factor is
$exp(-iE_{Re}t-\frac{\Gamma}{2} t)$ and the $\Gamma$ corresponds
to the total width and $E_{Re}-i\frac{\Gamma}{2}$ is a solution of
the Schr\"{o}dinger equation with a complex potential.

For the P-wave, the binding energy is positive and the angular
momentum barrier prevents dissociation of the bound state. It is
noted that since $\psi(0)=0$  for the P-wave, the imaginary part
of the complex potential which is proportional to $\delta({\bf
r})$, does not result in an imaginary part to the eigenenergy. The
dissolution mechanism of the bound state is the quantum
tunnelling. By the WKB approximation method \cite{Gasiorowicz},
the tunnelling transition probability is
$exp[-2\int_{a}^{b}\sqrt{2\mu(V-E)}dr]$, thus the total width of
the P-wave bound state would be $2\mu\hspace{0.05cm}
exp[-2\int_{a}^{b}\sqrt{2\mu (V-E)}dr]$, where
$\mu=\frac{m_{p}}{2}$ is the reduced mass.

Substituting the potential no matter real and complex, into the
Schr\"{o}dinger equation, and solving it, one obtain both the
eigenenergies and eigenfunctions of both S-wave and P-wave bound
states. Then we can evaluate the masses and total widths of the
bound states. That is the strategy of this work.

This paper is organized as follows. After this introduction, we
derive the formulation for the complex potential with a brief
introduction of the linear $\sigma$ model. In sec. III, we
substitute the potential into the Schr\"{o}dinger equation and
solve it to obtain the numerical result of the eigenenergy and
eigenfunction, then obtain the masses and widths of the S- and
P-wave bound states. In the section we also present all relevant
parameters. The last section is devoted to our conclusion and
discussion.\\

\section{The formulation}

(1) The necessary information about the model.

In the linear $\sigma$ model, the effective Lagrangian is
\begin{equation}
L=g\bar{\psi}(\sigma+\gamma_5{\mbox{\boldmath $
\tau$}}\cdot{\mbox{\boldmath $\pi$}})\psi,
\end{equation}
where $\psi$ is the wavefunction of the nucleon. When we calculate
the scattering amplitude, we introduce a form factor to compensate
the off-shell effects of the exchanged mesons. At each vertex, the
form factor is written as \cite{Liu}
\begin{equation}
\label{form} {\Lambda^2-M_m^2\over \Lambda^2-q^2},
\end{equation}
where $\Lambda$ is a phenomenological parameter and its value is
near 1 GeV. It is observed that as $q^2\to 0$ it becomes a
constant and if $\Lambda\gg M_m$, it turns to be unity. In the
case, as the distance is infinitely large, the vertex looks like a
perfect point, so the form factor is simply 1 or a constant.
Whereas, as $q^2\rightarrow \infty$, the form factor approaches to
zero, namely, in this situation, the distance becomes very small,
the inner structure (quark, gluon degrees of freedom) would
manifest itself and the whole picture of hadron interaction is no
longer valid, so the form factor is zero which cuts off the end
effects.

To derive an effective potential, one sets $q_0=0$ and writes down
the elastic scattering amplitude in the momentum space and then
carries out a Fourier transformation turning the amplitude into an
effective potential in the configuration space. Following the
standard procedure \cite{Landau}, we derive the effective
potential from the scattering amplitude. Below, we present some
details about the individual parts of the potential.\\

\noindent{(2) The effective potentials.}

(i) The real part of the potential.\\

Here we first consider the meson exchanges at the t-channel,
because the intermediate meson is a space-like, it cannot be on
its mass-shell, so that does not contribute to the imaginary part
of the effective potential. Then we will go on discussing the
s-channel contributions in subsection (ii).

a. Via exchanging $\pi$-meson:

The effective vertex is
\begin{eqnarray}
L &=& g\bar{\psi}\gamma_5{\mbox{\boldmath $\tau$}}\cdot
{\mbox{\boldmath $\pi$}}\psi,
\end{eqnarray}
and obviously only $\pi^{0}$ can be  exchanged  in our case.

The scattering amplitude in the momentum space is
 \begin{eqnarray}
  V_\pi(\mathbf{q})&=&-\frac{\displaystyle g_{NN\pi}^2}{\displaystyle 4m^2({\mathbf{q}}^2+m^2_\pi)}
 ({\mbox{\boldmath $\sigma$}}_1\cdot{\mathbf{q}})({\mbox{\boldmath $\sigma$}}_2\cdot{\mathbf{q}})
 \bigg(\frac{\displaystyle \Lambda^2-m^2_\pi}{\displaystyle
 \Lambda^2+{\mathbf{q}}^2}\bigg)^2.
\end{eqnarray}

Following the standard procedure, we carry out a Fourier
transformation on $V_\pi(\mathbf{q})$ and obtain the effective
potential in the configuration space:
 \begin{eqnarray}
   V_\pi(r)&=&\frac{\displaystyle g_{NN\pi}^2}{\displaystyle 4m^2}
   ({\mbox{\boldmath $\sigma$}}_1\cdot{\mbox{\boldmath $\nabla$}})
   ({\mbox{\boldmath $\sigma$}}_2\cdot{\mbox{\boldmath $\nabla$}})f_\pi(r)
\end{eqnarray}
where
 \begin{eqnarray}
  f_\pi(r)& = &\frac{\displaystyle e^{-m_\pi r}}{\displaystyle 4\pi r}
  -\frac{\displaystyle e^{-\Lambda r}}{\displaystyle 4\pi r}
   + \frac{\displaystyle(m_\pi^2-\Lambda^2)e^{-\Lambda r}}{\displaystyle
   8\pi\Lambda}.
   \end{eqnarray}

b. Via $\sigma$ and $\rho$ and $\omega$ exchanges.

The effective vertices are respectively
\begin{equation}
L_{\sigma}=g\bar\psi \psi\sigma,
\end{equation}
\begin{equation}
L_{\rho}=g_{_{NN}\rho}\bar\psi\gamma_{\mu}\tau^a\psi
A^{a\mu},\;\;\;\;\; a=1,2,3,
\end{equation}
\begin{equation}
L_{\omega}=g_{_{NN}\omega}\bar\psi\gamma_{\mu}\psi \omega^{\mu}.
\end{equation}

The scattering amplitude via exchanging $\sigma$-meson is
 \begin{eqnarray*}
  V_\sigma({\mathbf{q}})&=&\frac{\displaystyle g_{NN\sigma}^2}{\displaystyle 4m^2({\mathbf{q}}^2+m^2_\sigma)}
  \big[4m^2-4{\mathbf{p}}^2-{\mathbf{q}}^2 -4({\mathbf{p}}\cdot{\mathbf{q}})
  -i{\mbox{\boldmath $\sigma$}}\cdot({\mathbf{q}}\times{\mathbf{p}})\big]
\bigg(\frac{\displaystyle \Lambda^2-m_\sigma^2}{\displaystyle
\Lambda^2+{\mathbf{q}}^2}\bigg)^2,
 \end{eqnarray*}
through a Fourier transformation, the potential is
 \begin{eqnarray*}
 V_\sigma(r)=\frac{\displaystyle g_{NN\sigma}^2}{\displaystyle 4m^2}
 \bigg[4m^2f_\sigma(r)-4{\mathbf{p}}^2f_\sigma(r)+{\mbox{\boldmath$\nabla$}}^2f_\sigma(r)
 +4i({\mathbf{p}}\cdot{\mathbf{r}})F_\sigma(r)
-2({\mathbf{L}}\cdot{\mathbf{S}})F_\sigma(r)\bigg],
 \end{eqnarray*}
where
\begin{eqnarray*}
  f_\sigma(r)&=&\frac{\displaystyle e^{-m_\sigma r}}{\displaystyle 4\pi r}
  -\frac{\displaystyle e^{-\Lambda r}}{\displaystyle 4\pi r}
  +\frac{\displaystyle (m_\sigma^2-\Lambda^2)e^{-\Lambda r} }{\displaystyle 8\pi\Lambda}\\
  F_\sigma(r)&=&\frac{1}{\displaystyle r}\frac{\displaystyle \partial}{\displaystyle \partial
  r}f_\sigma(r).
   \end{eqnarray*}

Via exchanging vector-meson $\rho$(only $\rho^{0}$ contributes),
 the effective potential is
 \begin{eqnarray*}
  V_\rho(r)&=&\frac{\displaystyle g_{NN\rho}^2}{\displaystyle 4m^2}
  \bigg[4m^2f_\rho(r)-{\mbox{\boldmath$\nabla$}}^2f_\rho(r)+4{\mathbf{p}}^2f_\rho(r)
  +2({\mathbf{L}}\cdot {\mathbf{S}})F_\rho(r) -4i({\mathbf{p}}\cdot {\mathbf{r}})F_\rho(r) \\&&
  +({\mbox{\boldmath $\sigma$}}_1\cdot{\mbox{\boldmath $\nabla$}})
  ({\mbox{\boldmath $\sigma$}}_2\cdot{\mbox{\boldmath $\nabla$}})f_\rho(r)
\bigg]
  \end{eqnarray*}
where
 \begin{eqnarray*}
 f_\rho(r)&=&\frac{\displaystyle e^{-m_\rho r}}{\displaystyle 4\pi r}
 -\frac{\displaystyle e^{-\Lambda r}}{\displaystyle 4\pi r}
 +\frac{\displaystyle (m_\rho^2-\Lambda^2)e^{-\Lambda r} }{\displaystyle 8\pi\Lambda}\\
 F_\rho(r)&=&\frac{1}{\displaystyle r}\frac{\displaystyle \partial}{\displaystyle \partial
 r}f_\rho(r).
 \end{eqnarray*}

For exchanging an $\omega$ vector meson, the expression is similar
to that in the $\rho$ case, but has an opposite sign to the $\rho$
contribution due to the G-parity \cite{eberhard,Richard}, thus one
only needs to replace the corresponding parameter values, such as
the mass and coupling constant for $\rho$ by that for $\omega$ and
add a minus sign in front of all the terms of $V_\rho(r)$. For
saving space, we dismiss the concrete expression for $\omega$
exchange.

c. The real part of the potential

A synthesis of all the individual contributions derived above
stands as the real part of the effective potential, namely the
traditional part of the effective potential as
 \begin{eqnarray*}
  V_{eff}(r)&=&V_\pi(r)+V_\sigma(r)+V_\rho(r)+V_\omega(r) \\
 &=&V_{0}(r)+V_{LS}(r)+V_{pet}(r)+V_{T}
 +V_{SS}.
 \end{eqnarray*}
In the expression the leading part of the potential is
 \begin{eqnarray*}
V_{0}&=&{\bigg(} \frac{\displaystyle e^{-\Lambda r}\Lambda
g_{NN\sigma}^2}{\displaystyle 8\pi}
 - \frac{\displaystyle e^{-\Lambda r}m_\sigma^2g_{NN\sigma}^2}{\displaystyle 8\pi\Lambda}
 + \frac{\displaystyle e^{-\Lambda r}g_{NN\sigma}^2}{\displaystyle 4\pi r}
 - \frac{\displaystyle e^{-m_\sigma r}g_{NN\sigma}^2}{\displaystyle 4\pi r}{\bigg)}
  + {\bigg(}-\frac{\displaystyle e^{-\Lambda r}\Lambda g_{NN\rho}^2}{\displaystyle 8\pi} \\&&
+\frac{\displaystyle e^{-\Lambda r}m_\rho^2
g_{NN\rho}^2}{\displaystyle 8\pi\Lambda}
 -\frac{\displaystyle e^{-\Lambda r}g_{NN\rho}^2}{\displaystyle 4\pi r}
 +\frac{\displaystyle e^{-m_\rho r}g_{NN\rho}^2}{\displaystyle 4\pi r}{\bigg)}
 +{\bigg(}\frac{\displaystyle e^{-\Lambda r}\Lambda g_{NN\omega}^2}{\displaystyle 8\pi}
  -\frac{\displaystyle e^{-\Lambda r}m_\omega^2 g_{NN\omega}^2}{\displaystyle 8\pi\Lambda}\\&&
   +\frac{\displaystyle e^{-\Lambda r}g_{NN\omega}^2}{\displaystyle 4\pi r}
   -\frac{\displaystyle e^{-m_\omega r}g_{NN\omega}^2}{\displaystyle 4\pi
   r}{\bigg)}.
\end{eqnarray*}
The spin-orbit term is
\begin{eqnarray*}
 V_{LS}&=&{\Bigg[}\bigg(\frac{\displaystyle e^{-\Lambda r}g_{NN\sigma}^2}{\displaystyle 8m^2\pi r^3}
 -\frac{\displaystyle e^{-m_\sigma r}g_{NN\sigma}^2}{\displaystyle 8m^2\pi r^3}
 +\frac{\displaystyle e^{-\Lambda r}\Lambda g_{NN\sigma}^2}{\displaystyle 8m^2\pi r^2}
 -\frac{\displaystyle e^{-m_\sigma r}g_{NN\sigma}^2m_\sigma}{\displaystyle 8m^2\pi r^2}
 +\frac{\displaystyle e^{-\Lambda r}\Lambda^2g_{NN\sigma}^2}{\displaystyle 16m^2\pi r}  \\&&
-\frac{\displaystyle e^{-\Lambda
r}m_\sigma^2g_{NN\sigma}^2}{\displaystyle 16m^2\pi r}\bigg) +
\bigg(\frac{\displaystyle e^{-\Lambda
r}g_{NN\rho}^2}{\displaystyle8m^2\pi r^3} -\frac{\displaystyle
e^{-m_\rho r}g_{NN\rho}^2}{\displaystyle8m^2\pi r^3}
+\frac{\displaystyle e^{-\Lambda r}\Lambda
g_{NN\rho}^2}{\displaystyle 8m^2\pi r^2} -\frac{\displaystyle
e^{-m_\rho r}m_\rho g_{NN\rho}^2}{\displaystyle 8m^2\pi r^2} \\&&
+\frac{\displaystyle e^{-\Lambda r}\Lambda^2
g_{NN\rho}^2}{\displaystyle 16m^2\pi r}
 -\frac{\displaystyle e^{-\Lambda r}m_\rho^2 g_{NN\rho}^2}{\displaystyle 16m^2\pi r}\bigg)
 + \bigg(-\frac{\displaystyle e^{-\Lambda r}g_{NN\omega}^2}{\displaystyle8m^2\pi r^3}
 +\frac{\displaystyle e^{-m_\omega r}g_{NN\omega}^2}{\displaystyle8m^2\pi r^3}
-\frac{\displaystyle e^{-\Lambda r}\Lambda
g_{NN\omega}^2}{\displaystyle 8m^2\pi r^2} \\&&
+\frac{\displaystyle e^{-m_\omega r}m_\omega
g_{NN\omega}^2}{\displaystyle 8m^2\pi r^2} - \frac{\displaystyle
e^{-\Lambda r}\Lambda^2 g_{NN\omega}^2}{\displaystyle16m^2\pi r}
+\frac{\displaystyle e^{-\Lambda r}m_\omega^2
g_{NN\omega}^2}{\displaystyle 16m^2\pi r}
\bigg)\Bigg](\mathbf{L}\cdot\mathbf{S}).
\end{eqnarray*}
The relativistic correction which in our later numerical
computations is treated as a perturbation to the leading part, is
 \begin{eqnarray*}
 V_{pet}&=&\bigg(\frac{ \displaystyle e^{-\Lambda r}\Lambda^3g_{NN\sigma}^2}{\displaystyle 8m^2\pi}
 -\frac{ \displaystyle e^{-\Lambda r}\Lambda m_\sigma^2 g_{NN\sigma}^2}{\displaystyle 8m^2\pi}
 +\frac{\displaystyle e^{-\Lambda r}g_{NN\sigma}^2}{\displaystyle 4m^2\pi r^3}
 -\frac{\displaystyle e^{-m_\sigma r}g_{NN\sigma}^2}{\displaystyle 4m^2\pi r^3}
 +\frac{\displaystyle e^{-\Lambda r}\Lambda g_{NN\sigma}^2}{\displaystyle 4m^2\pi r^2} \\&&
-\frac{\displaystyle e^{-m_\sigma r}m_\sigma
g_{NN\sigma}^2}{\displaystyle 4m^2\pi r^2} +\frac{\displaystyle
e^{-\Lambda r}\Lambda^2g_{NN\sigma}^2}{\displaystyle 8m^2\pi r}
+\frac{ \displaystyle e^{-\Lambda r} m_\sigma^2
g_{NN\sigma}^2}{\displaystyle 8m^2\pi r} -\frac{ \displaystyle
e^{-m_\sigma r} m_\sigma^2 g_{NN\sigma}^2}{\displaystyle 4m^2\pi
r}\bigg)\\&& +\bigg(\frac{\displaystyle 3e^{-\Lambda
r}\Lambda^3g_{NN\rho}^2}{\displaystyle 32m^2\pi} -\frac{
\displaystyle 3e^{-\Lambda r}\Lambda m_\rho^2
g_{NN\rho}^2}{\displaystyle 32m^2\pi} +\frac{\displaystyle
e^{-\Lambda r}g_{NN\rho}^2}{\displaystyle 4m^2\pi r^3}
-\frac{\displaystyle e^{-m_\rho r}g_{NN\rho}^2}{\displaystyle
4m^2\pi r^3} +\frac{\displaystyle e^{-\Lambda
r}g_{NN\rho}^2\Lambda}{\displaystyle 4m^2\pi r^2} \\&&
-\frac{\displaystyle e^{-m_\rho
r}g_{NN\rho}^2m_\rho}{\displaystyle 4m^2\pi r^2} +\frac{
\displaystyle e^{-\Lambda r}\Lambda^2g_{NN\rho}^2}{\displaystyle
8m^2\pi r} + \frac{\displaystyle e^{-\Lambda r} m_\rho^2
g_{NN\rho}^2}{\displaystyle 16m^2\pi r} -\frac{ \displaystyle
3e^{-m_\rho r} m_\rho^2 g_{NN\rho}^2}{\displaystyle 16m^2\pi
r}\bigg)\\&& +\bigg(-\frac{ 3\displaystyle e^{-\Lambda
r}\Lambda^3g_{NN\omega}^2}{\displaystyle 32m^2\pi}
+\frac{\displaystyle 3e^{-\Lambda r}\Lambda m_\omega^2
g_{NN\omega}^2}{\displaystyle32m^2\pi} -\frac{\displaystyle
e^{-\Lambda r}g_{NN\omega}^2}{\displaystyle 4m^2\pi r^3}
+\frac{\displaystyle e^{-m_\omega r}g_{NN\omega}^2}{\displaystyle
4m^2\pi r^3} -\frac{\displaystyle e^{-\Lambda
r}g_{NN\omega}^2\Lambda}{\displaystyle 4m^2\pi r^2}  \\&&
+\frac{\displaystyle e^{-m_\omega
r}g_{NN\omega}^2m_\omega}{\displaystyle 4m^2\pi r^2} -\frac{
\displaystyle e^{-\Lambda r}\Lambda^2g_{NN\omega}^2}{\displaystyle
8m^2\pi r} -\frac{\displaystyle e^{-\Lambda r} m_\omega^2
g_{NN\omega}^2}{\displaystyle 16m^2\pi r} +\frac{ \displaystyle
3e^{-m_\omega r} m_\omega^2 g_{NN\omega}^2}{\displaystyle 16m^2\pi
r}\bigg).
\end{eqnarray*}
The tensor potential is
  \begin{eqnarray*}
  V_{T} &=&\bigg( \frac{\displaystyle e^{-\Lambda r}\Lambda^3 g_{NN\pi}^2}{\displaystyle 96m^2\pi}
  -\frac{\displaystyle e^{-\Lambda r}\Lambda m_\pi^2g_{NN\pi}^2}{\displaystyle 96m^2\pi}
  +\frac{\displaystyle e^{-\Lambda r}g^2}{16m^2\pi r^3}
  -\frac{\displaystyle e^{-m_\pi r}g_{NN\pi}^2}{16m^2\pi r^3}
  +\frac{\displaystyle e^{-\Lambda r}\Lambda g_{NN\pi}^2}{16m^2\pi r^2} r \\ &&
 -\frac{\displaystyle e^{-m_\pi r}m_\pi g_{NN\pi}^2}{16m^2\pi r^2}
 +\frac{\displaystyle e^{-\Lambda r}\Lambda^2 g_{NN\pi}^2}{32m^2\pi r}
 -\frac{\displaystyle e^{-\Lambda r}m_\pi^2 g_{NN\pi}^2}{96m^2\pi r}
 -\frac{\displaystyle e^{-m_\pi r}m_\pi^2 g_{NN\pi}^2}{48m^2\pi r}\bigg) \\ &&
+\bigg(- \frac{\displaystyle e^{-\Lambda r}\Lambda^3
g_{NN\rho}^2}{\displaystyle 96m^2\pi} +\frac{\displaystyle
e^{-\Lambda r}\Lambda m_\rho^2g_{NN\rho}^2}{\displaystyle
96m^2\pi} -\frac{\displaystyle e^{-\Lambda
r}g_{NN\rho}^2}{16m^2\pi r^3} +\frac{\displaystyle e^{-m_\rho
r}g_{NN\rho}^2}{16m^2\pi r^3} -\frac{\displaystyle e^{-\Lambda
r}\Lambda g_{NN\rho}^2}{16m^2\pi r^2}  \\ && +\frac{\displaystyle
e^{-m_\rho r}m_\rho g_{NN\rho}^2}{16m^2\pi r^2}
-\frac{\displaystyle e^{-\Lambda r}\Lambda^2
g_{NN\rho}^2}{32m^2\pi r} +\frac{\displaystyle e^{-\Lambda
r}m_\rho^2 g_{NN\rho}^2}{96m^2\pi r} +\frac{\displaystyle
e^{-m_\rho r}m_\rho^2 g_{NN\rho}^2}{48m^2\pi r}\bigg) \\&& +
\bigg(\frac{\displaystyle e^{-\Lambda
r}\Lambda^3g_{NN\omega}^2}{\displaystyle 96m^2\pi}
 -\frac{\displaystyle e^{-\Lambda r}\Lambda m_\omega^2g_{NN\omega}^{2}}{\displaystyle 96m^2\pi}
 +\frac{\displaystyle e^{-\Lambda r}g_{NN\omega}^2}{16m^2\pi r^3}
 -\frac{\displaystyle e^{-m_\omega r}g_{NN\omega}^2}{16m^2\pi r^3} \\&&
 +\frac{\displaystyle e^{-\Lambda r}\Lambda g_{NN\omega}^2}{16m^2\pi r^2}
 -\frac{\displaystyle e^{-m_\omega r}m_\omega g_{NN\omega}^2}{16m^2\pi r^2}
+\frac{\displaystyle e^{-\Lambda r}\Lambda^2
g_{NN\omega}^2}{32m^2\pi r} -\frac{\displaystyle e^{-\Lambda
r}m_\omega^2 g_{NN\omega}^2}{96m^2\pi r} \\&& -\frac{\displaystyle
e^{-m_\omega r}m_\omega^2 g_{NN\omega}^2}{48m^2\pi r}\bigg)
\bigg[\frac{3({\mbox{\boldmath
$\sigma$}}_1\cdot\mathbf{r})({\mbox{\boldmath
$\sigma$}}_2\cdot\mathbf{r})}{r^2} -({\mbox{\boldmath
$\sigma$}}_1\cdot{\mbox{\boldmath $\sigma$}}_2)\bigg],
\end{eqnarray*}
and the spin-spin term is
\begin{eqnarray*}
V_{SS}&=& \bigg(\frac{\displaystyle e^{-\Lambda
r}\Lambda^3g_{NN\pi}^2}{\displaystyle 96m^2\pi}
-\frac{\displaystyle e^{-\Lambda r}\Lambda
m_\pi^2g_{NN\pi}^2}{\displaystyle 96m^2\pi} +\frac{\displaystyle
e^{-\Lambda r}m_\pi^2g_{NN\pi}^2}{\displaystyle 48m^2\pi r}
-\frac{\displaystyle e^{-m_\pi r}m_\pi^2g_{NN\pi}^2}{\displaystyle
48m^2\pi r}\bigg) \\&& +\bigg(\frac{\displaystyle e^{-\Lambda
r}\Lambda^3g_{NN\rho}^2}{\displaystyle 48m^2\pi}
-\frac{\displaystyle e^{-\Lambda r}\Lambda
m_\rho^2g_{NN\rho}^2}{\displaystyle 48m^2\pi} +\frac{\displaystyle
e^{-\Lambda r}m_\rho^2g_{NN\rho}^2}{\displaystyle 24m^2\pi r}
-\frac{\displaystyle e^{-m_\rho
r}m_\rho^2g_{NN\rho}^2}{\displaystyle 24m^2\pi r}\bigg)\\&&
+\bigg(-\frac{\displaystyle e^{-\Lambda
r}\Lambda^3g_{NN\omega}^2}{\displaystyle 48m^2\pi}
 +\frac{\displaystyle e^{-\Lambda r}\Lambda m_\omega^2g_{NN\omega}^2}{\displaystyle 48m^2\pi}
 -\frac{\displaystyle e^{-\Lambda r}m_\omega^2g_{NN\omega}^2}{\displaystyle 24m^2\pi r}
 \nonumber\\
 &&+\frac{\displaystyle e^{-m_\omega r}m_\omega^2g_{NN\omega}^2}{\displaystyle 24m^2\pi r}\bigg)
 ({\mbox{\boldmath $\sigma$}}_{1}\cdot{\mbox{\boldmath
 $\sigma$}}_{2}).
\end{eqnarray*}

d. The case of $p\bar p$ is different from the deuteron where the
constituents are $p$ and $n$, namely there is a $p \bar p$
annihilation at the s-channel, which would contribute a delta
function to the effective potential.

If $p\bar p$ is in S-wave with quantum number $I^{G}
J^{PC}~=~0^{+}(0^{-+}$), in the s-channel only a $0^{+}(0^{-+})$
meson can be exchanged. Here we only consider the lowest-lying
pseudoscalar mesons of $0^{-+}$ $\eta$ and $\eta '$. Their
contribution can be written as
\begin{eqnarray*}
   V_{\eta}^{'}(r)&=& \frac{\displaystyle g_{NN\eta}^2}{\displaystyle(4m^2-m^2_\eta)}
 \bigg(\frac{\displaystyle \Lambda^2-m^2_\eta}{\displaystyle  4m^2-\Lambda^2}\bigg)^2
 \bigg[-1+\frac{\displaystyle({\mbox{\boldmath $\sigma$}}_1\cdot{\mbox{\boldmath$\nabla$}})
 ({\mbox{\boldmath $\sigma$}}_2\cdot{\mbox{\boldmath$\nabla$}})}{\displaystyle 2m^2}
 -\frac{{\mbox{\boldmath{$\nabla$}}}^2}{2m^2}\bigg]\delta^3({\mathbf
 r}),
 \end{eqnarray*}
where $m$ is the invariant mass $\sqrt{(p_1+p_2)^2}$ and
$p_1,\;p_2$ are the four-momenta of the constituents $p$ and $\bar
p$ respectively, and it is very close to $2m_p$. For the
contribution of $\eta '$, one only needs to replace the
corresponding parameters values. It is noted that these
contributions still belong to the real part of the effective
potential. Below, we will derive the imaginary contributions
induced by the absorptive part of loops at s-channel.

(ii) The imaginary part of the complex potential.\\

The corresponding Feynman diagrams are shown in Figs.1 and 2.
Fig.1 is the self-energy of $\eta$ and $\eta '$ which are
off-shell and Fig.2 is a box diagram. Obviously, the elastic
scattering of $p\bar p$ is a strong-interaction process, so that
parity, isospin etc. quantum numbers must be conserved and as long
as the $p\bar p$ bound state is of the $0^{-+}$ structure, only
$\eta$ and $\eta'$ can be exchanged in the s-channel (we neglect
higher-resonances).

The concerned couplings are \cite{Harada,Achasov,lucio}
\begin{eqnarray}
L_{PP\sigma} &=&
-\frac{\gamma_{PP \sigma }}{\sqrt{2}}\sigma\partial_{\mu}{\mbox{\boldmath $P$}}\cdot\partial_{\mu}{\mbox{\boldmath $P$}},\\
L_{VVp} &=& g_{VVp}\varepsilon^{\mu\nu\lambda\sigma}\partial_{\mu}
V_{\nu}({\mbox{\boldmath $P$}}\cdot\partial_{\lambda}
{\mbox{\boldmath $V$}}_{\sigma}),
\end{eqnarray}
here ${\bf P}$ stands as pseudoscalar mesons, such as $\pi$,
$\eta$ and $\eta '$ etc.  and  ${\bf V}$ denotes vector mesons,
such as $\omega$ and $\rho$ etc.

The imaginary part of the potential is obtained in the following
way. First, we calculate the absorptive part of the loops by the
Cutkosky cutting rule in the momentum space \cite{Itsykson} and
carry out a Fourier transformation turning it into an imaginary
part of the complex potential.

(i) The contribution induced by the self-energy of $\eta$ and
$\eta'$.

We have obtained
\begin{eqnarray*}
   V_{{\mathbf{Im}}_{1}}(r) &=&-\frac{\displaystyle \gamma_{\eta\eta\sigma}^2g_{NN\eta}^2\displaystyle (4m^2-m_{\sigma}^2+m_{\eta}^2)^2}{\displaystyle 512\pi m^2\displaystyle (4m^2-m^2_\eta)^2}
    \sqrt{-16m^2 m_{\eta}^2+(4m^2-m_{\sigma}^2+m_{\eta}^2)^2}, \\&&
    \times\bigg(\frac{\displaystyle \Lambda^2-m^2_\eta}{\displaystyle
    4m^2-\Lambda^2}\bigg)^2\delta^3(\textbf{r}), \;\;\;\;\;\;{\rm for\; Fig.1(a)}.
 \end{eqnarray*}

\begin{eqnarray*}
   V_{{\mathbf{Im}}_{2}}(r)&=&-\frac{\displaystyle \gamma_{\eta\eta f_{0}}^2g_{NN\eta}^2\displaystyle (4m^2-m_{f_{0}}^2+m_{\eta}^2)^2}{\displaystyle 512\pi m^2\displaystyle (4m^2-m^2_\eta)^2}
    \sqrt{-16m^2 m_{\eta}^2+(4m^2-m_{f_{0}}^2+m_{\eta}^2)^2}\\&&
   \times \bigg(\frac{\displaystyle \Lambda^2-m^2_\eta}{\displaystyle
    4m^2-\Lambda^2}\bigg)^2\delta^3(\textbf{r}), \;\;\;\;\;\;{\rm for\; Fig.1(b)}.
 \end{eqnarray*}
\begin{eqnarray*}
   V_{{\mathbf{Im}}_{3}}(r)&=&-\frac{\displaystyle \gamma_{\pi\eta a_{0}}^2g_{NN\eta}^2\displaystyle (4m^2-m_{a_{0}}^2+m_{\pi}^2)^2}{\displaystyle 512\pi m^2\displaystyle (4m^2-m^2_\eta)^2}
    \sqrt{-16m^2 m_{\pi}^2+(4m^2-m_{a_{0}}^2+m_{\pi}^2)^2}\\&&
   \times \bigg(\frac{\displaystyle \Lambda^2-m^2_\eta}{\displaystyle
    4m^2-\Lambda^2}\bigg)^2\delta^3(\textbf{r}), \;\;\;\;\;\;{\rm for\; Fig.1(c)}.
 \end{eqnarray*}
and
\begin{eqnarray*}
V_{{\mathbf{Im}}_{4}}(r)&=&-\frac{\displaystyle
g_{\eta\rho\rho}^{2}g_{NN\eta}^{2}(16m^{4}-4m^{2}m_{\rho}^{2
})}{\displaystyle 128\pi m^{2}(4m^{2}-m_{\eta}^{2})^{2}}
\sqrt{16m^4-16m^2 m_{\rho}^2}
    \bigg(\frac{\displaystyle \Lambda^2-m^2_\eta}{\displaystyle
    4m^2-\Lambda^2}\bigg)^2\delta^3(\textbf{r}), \;\;\;\;\;\;{\rm for\; Fig.1(d)}.
\end{eqnarray*}

b. The contributions induced by the box diagram
\begin{eqnarray*}
V_{{\mathbf{Im}}_{box}}({\bf
p})&=&2g_{NN\sigma}^{2}g_{NN\eta}^{2}\int\frac{d^{4}q}{(2\pi)^4}\overline{v}(p_{2})\frac{p\!\!\!\slash_{1}
-q\!\!\!\slash+m}{(p_{1}-q)^{2}-m^2}\gamma^{5}u(p_{1})\overline{u}(p_{3})\gamma^{5}
\frac{p\!\!\!\slash_{3}-q\!\!\!\slash+m}{(p_{3}-q)^2-m^2}v(p_{4})\\&&
\times(i\pi)^2\delta(q^2-m_{\eta}^2)\delta([p_{1}+p_{2}-q]^{2}-m_{\sigma}^2)\Big(\frac{\Lambda^2-m^2}{(p_{1}-q)^{2}-\Lambda^2}\Big)^2
\Big(\frac{\Lambda^2-m^2}{(p_{3}-q)^{2}-\Lambda^2}\Big)^2.
\end{eqnarray*}

Taking the Fourier transformation, we obtain
\begin{eqnarray*}
V_{{\mathbf{Im}}_{box}}(r)
&=&\int^{\pi}_{0}\Bigg\{\frac{g_{NN\sigma}^2g_{NN\eta}^{2}{\mathcal{B}}
(\Lambda^{2}-m^2)^{4}}{16\pi
{\mathcal{A}}[{\mathcal{A}}^2-{\mathcal{B}}^2-2({\mathcal{A}}\sqrt{m^{2}+{\mathbf{p}}^{2}}
-{\mathcal{B}}{\mathbf{p}}\cos(\theta))]^{2}}\\&&
\times\frac{[-m^{2}-(m+\frac{{\mathbf{p}}^{2}}{2m}-{\mathcal{A}})^{2}
+2m(m+\frac{{\mathbf{p}}^2}{2m}-{\mathcal{A}})]Sin(\theta)}{[m^2-{\mathcal{B}}^{2}
+{\mathcal{A}}^{2}-\Lambda^{2}-2({\mathcal{A}}\sqrt{m^2+{\mathbf{p}}^{2}}-{\mathcal{B}}{\mathbf{p}}
\cos(\theta))]^{4}}\Bigg\}d{\theta}\delta^{3}({\mathbf{r}}),
\end{eqnarray*}
where
\begin{eqnarray*}
{\mathcal{A}}&=&\sqrt{m_\eta^2+{\mathcal{B}}^2},
\end{eqnarray*}
and
\begin{eqnarray*}
{\mathcal{B}}&=&\frac{1}{4}\sqrt{16m^2-8m_{\eta}^{2}-8m_{\sigma}^{2}
+16{\mathbf{p}}^{2}+\frac{m_{\eta}^{4}}{m^{2}+{\mathbf{p}}^{2}}
-\frac{2m_{\eta}^2m_{\sigma}^2}{m^{2}+{\mathbf{p}}^2}+\frac{m_{\sigma}^{4}}{m^{2}+{\mathbf{p}}^{2}}}.
\end{eqnarray*}

Expanding the expression with respect to $|\mathbf{p}|$ and
keeping terms up to ${\bf p}^{2}$, we have the final expression as
\begin{eqnarray*}
V_{{\mathbf{Im}}_{box}}(r)&=&\Big[\frac{-g_{NN\sigma}^{2}g_{NN\eta}^{2}(4m^2-m_{\sigma}^{2}+m_{\eta}^{2})^{2}(m^2-\Lambda^2)^4}
{16\pi m^4(-4m^2+m_{\sigma}^{2}+m_{\eta}^{2})^{2}(-2m^2+m_{\sigma}^{2}+m_{\eta}^{2}-2\Lambda^{2})^{4}} \\
&&\times\sqrt{16m^{4}+(m_{\eta}^{2}-m_{\sigma}^{2})^{2}-8m^{2}(m_{\eta}^{2}+m_{\sigma}^{2})^{2}}+{\mathcal{O}}(m,m_{\sigma},m_{\eta},
,\Lambda){\mathbf{p}}^{2}\Big]\delta^{3}({\mathbf{r}})
\end{eqnarray*}
where the coefficient
${\mathcal{O}}(m,m_{\sigma},m_{\eta},\Lambda)$ of ${\bf p}^2$ is
obtained by a tedious but straightforward calculation, here for
saving space we ignore the details.

Finally we obtain the imaginary part of the potential as
\begin{eqnarray*}
V_{{\mathbf{Im}}}(r)=V_{{\mathbf{Im}}_{1}}(r)+V_{{\mathbf{Im}}_{2}}(r)+V_{{\mathbf
{Im}}_{3}}(r)+V_{{\mathbf
{Im}}_{4}}(r)+V_{{\mathbf{Im}}_{box}}(r).
\end{eqnarray*}

It is also noted that for the P-wave resonance, the wavefunction
at origin is zero, i.e. $\psi(0)=0$, at the leading order there is
no s-channel contribution to the effective potential, and neither
the imaginary part in the final effective potential, since in our
approximation, all of them are proportional to $\delta^3({\bf r})$.\\

In the practical computation, a popular approximation \cite{Lucha}
for the delta function
\begin{eqnarray*}
\delta^3(\mathbf{r})\propto\frac{\alpha^{3}}{\pi^{\frac{3}{2}}}e^{-\alpha^{2}r^{2}}
\end{eqnarray*}
is adopted.

\section{Numerical results}

\hspace{6mm}By solving the Schr\"{o}dinger equation, we obtain the
zero-th order engenenergy and wavefunction, where the L-S coupling
and tensor terms are taken as perturbations and the imaginary part
of the complex potential is treated in two ways. In terms of the
traditional method of Quantum Mechanics, we can calculate the
corrections.

For the S-wave resonance we obtain the binding energy and the
total width as
\begin{table}[htb]
\begin{center}
\begin{tabular}{|c|c|c|c|c|c|c|c|c|c|} \hline
$m_{\sigma}$(GeV)&0.47&0.48&0.49&0.50&0.51&0.52&0.53&0.54&0.55\\
\hline $\Lambda$(GeV)&0.59&0.60&0.61&0.62&0.63&0.64&0.65&0.66&0.67\\
\hline $Es$(MeV)&-18.38&-17.23&-16.19&-15.30&-14.47&-13.67&-13.03&-12.42&-11.88\\
\hline $\Gamma_{s}$(MeV)&33.60&30.08&26.94&24.15&21.67&19.49&17.57&15.90&14.45\\
\hline
\end{tabular}
\end{center}
\caption{{the theoretical results for the S-wave with perturbative
method }}
\end{table}

\vspace {0.3cm}

The number listed in table.1 are obtained in terms of the
perturbation method. Namely, we take the imaginary part of the
complex potential as a perturbation as well as the L-S coupling
and tensor terms,
$$\Delta E=<\psi_0|\Delta H|\psi_0>,$$
where $\Delta H=\Delta H_{real}+i\Delta H_{imag}$ and $\psi_0$ is
the wavefunction of zero-th order.

As we sandwich the imaginary part of the complex potential between
$\psi_0$, the expectation value is the imaginary part of the
complex eigenenergy as
$$i\Delta E_{imag}=-i {\Gamma\over 2}=<\psi_0|iV_{\bf Im}(r)|\psi_0>,$$
and
$$iV_{\bf Im}(r)\equiv i\Delta H_{imag}.$$
Thus the total eigenenergy is
$$E=E_0+\Delta E_{real}+i\Delta E_{imag}=E_0+\Delta
E_{real}-i{\Gamma\over 2}.$$

Instead, one can solve the Schr\"odinger equation with a complex
potential which would be divided into two coupled differential
equations. The coupled equations cannot, in general, be solved
analytically, but only numerically. We obtain the complex
eigenenergy by solving the equation group and the results are
listed in table 2.

\begin{table}[htb]
\begin{center}
\begin{tabular}{|c|c|c|c|c|c|c|c|c|c|} \hline
$m_{\sigma}$(GeV)&0.47&0.48&0.49&0.50&0.51&0.52&0.53&0.54&0.55\\
\hline $\Lambda$(GeV)&0.59&0.60&0.61&0.62&0.63&0.64&0.65&0.66&0.67\\
\hline $Es$(MeV)&-14.10&-13.64&-13.17&-12.71&-12.25&-11.81&-11.38&-10.99&-10.62\\
\hline $\Gamma_{s}$(MeV)&21.90&19.48&17.29&15.29&13.48&11.87&10.45&9.19&8.11\\
\hline
\end{tabular}
\end{center}
\caption{{the calculated results for the S-wave with direct
program calculating  }}
\end{table}

\vspace{0.6cm}

Comparing the results in table 1 and 2, we find that there are
some deviations of no more than 30\%, and generally, the numbers
obtained in the perturbation method are a bit greater than them by
directly solving the coupled equations, but qualitatively, the two
sets are consistent. Considering the experimental errors and
theoretical uncertainties, we would conclude that the two sets of
numbers agree with each other.

For the P-wave, by solving the Schr\"{o}dinger equation, we obtain
the eigenenergy, and then by the WKB method, we can estimate the
dissociation rate, which turns out to be the width of the P-wave
resonance. The results are shown in Table 2.
\begin{table}[htb]
\begin{center}
\begin{tabular}{|c|c|c|c|c|c|c|c|c|c|} \hline
$m_{\sigma}$(GeV)&0.47&0.48&0.49&0.50&0.51&0.52&0.53&0.54&0.55\\
\hline $\Lambda$($\times 10^{-1}$ GeV)&8.25&8.51&8.76&9.02&9.28&9.54&9.81&10.08&10.35\\
\hline $Ep$(MeV)&0.35&0.43&0.54&0.63&0.72&0.80&0.87&0.94&1.08\\
\hline
$\Gamma_{p}$(MeV)&9.11&11.42&15.00&17.96&20.22&22.64&24.26&25.91&29.98\\
\hline
\end{tabular}
\end{center}
\caption{The calculated values for P-wave}
\end{table}
\vspace{0.3cm}

\vspace{0.6cm}

For the theoretical calculations, we have employed the following
parameters as inputs: $m=0.938$(GeV);\ $m_{\pi}=0.138$(GeV);\
$m_{\rho}=0.77$(GeV);\ $m_{\omega}=0.783$(GeV);\
$m_{\eta}=0.547$(GeV);\ $m_{\eta^{'}}=0.958$(GeV);\
$m_{f_{0}}=0.98$(GeV);\ $m_{a_{0}}=0.98$(GeV) \cite{PDG};\
$g_{NN\pi}=g_{NN\sigma}=13.5$;\ $g_{NN\rho}=g_{NN\omega}=3.25$
\cite{Lin};\ $\frac{g_{NN\eta}^2}{4\pi}=0.4$;\
$\frac{g_{NN\eta^{'}}^2}{4\pi}=0.6$ \cite{tiator};\
$\gamma_{\eta\eta\sigma}=4.11$(GeV$^{-1}$);\
$\gamma_{\eta\eta^{'}\sigma}=2.65$(GeV$^{-1}$);\ $\gamma_{\eta\eta
f_{0}}=1.72$(GeV$^{-1}$);\ $\gamma_{\eta\eta^{'}
f_{0}}=-9.01$(GeV$^{-1}$);\ $\gamma_{\pi\eta
a_{0}}=-6.80$(GeV$^{-1}$);\ $ \gamma_{\pi\eta^{'}
a_{0}}=-7.80$(GeV$^{-1}$) \cite{Black};\ $ g_{\eta\rho\rho
}=-16$(GeV$^{-1}$)\cite{lucio}\cite{Lublinsky}.

\section{Conclusion and discussion}

\hspace{6mm}In this work, in terms of the linear $\sigma$ model we
investigate the spectrum and total width of the possible
$p\bar{p}$ bound states. We consider two possibilities that the
observed enhancement is due to a $p\bar{p}$ bound state in S- or
P-waves respectively.

All the parameters employed in the calculations were obtained by
fitting the data of deuteron. With the very precise measurement on
the binding energy of deuteron and more or less accurate estimate
of the s-d mixing and charge radius, there is only a narrow window
in the parameter space \cite{Liu}. Namely there is almost not much
free room to adjust them, and neither is a large range for
changing our theoretical calculations, as long as the model is
employed. Therefore the newly observed resonance, if it is
experimentally confirmed, can also provide an opportunity to
further testify the linear $\sigma$ model.

We derive the effective potential between $p$ and $\bar{p}$, and
for the S-wave structure, we simply substitute it into the
schr\"{o}dinger equation to obtain the binding energy. We have
also calculated the absorptive part of the concerned loops by the
Cutkosky cutting rule and it becomes the imaginary part of the
potential. Differently from the deuteron case which is a bound
state of $pn$, for the $p\bar p$ case, there exist s-channel
processes (see the figures in the text), which can contribute a
real part (the tree level meson exchange) to the potential and an
imaginary part through the loop diagrams in the channel and both
of them are proportional to $\delta({\bf r})$ at the concerned
non-relativistic approximation. In this work, we ignore the
dispersive part of the loops because it depends on the
renormalization scheme and only makes a correction to the leading
contribution of the real part of the potential, but keep the
absorptive part which is the only source of the imaginary part of
the potential.

When we solve the Schr\"{o}dinger equation with a complex
potential, we have taken certain approximations to simplify the
calculations. Then we obtain the mass and total width of the
S-wave resonance.

It is also noted that due to the G-parity structure of the $NN$
and $N\bar N$ systems where $N$ refers to nucleons, the potentials
contributed by $\sigma$ and $\rho$ are of the same sign for the
$NN$ and $N\bar N$ systems, but the potential induced by $\pi^0$
and $\omega$ should have opposite signs for the two systems
\cite{eberhard,Richard}. In fact, for the deuteron case, which is
in the $pn$ structure, the contribution of $\pi^0$ is repulsive,
and so is that from $\rho$ and $\omega$. But for the $p\bar p$
system, $\pi^0$ and $\omega$ induce attractive potentials while
the contribution induced by $\sigma$ and $\rho$ remain unchanged.
It can qualitatively explain why the the binding energy for $p\bar
p$ (about $-18$ MeV) is more negative than that for deuteron
(about $-2.22$ MeV).

For the P-wave case, where the wave function of origin is zero,
i.e. $\psi(0)=0$, one does not need to calculate the s-channel
contribution. The angular momentum barrier prevents dissociation
of the bound state, but the quantum tunnelling leads to a final
dissolution of the bound state and this tunnelling rate determines
the total width or lifetime of the P-wave bound state of
$p\bar{p}$. Fig.3 shows the effective potential for S-wave and
P-wave respectively. The repulsive part at the region of small $r$
is due to the vector-meson exchange. One also notes that for the
S-wave, besides the repulsive core for small $r$ the potential is
attractive, whereas for the P-wave, there exists an angular
barrier which results in an positive binding energy, i.e. the
total mass of the P-wave resonance is above the threshold of
$2m_p$. Since the barrier is not high, the binding energy is not
far above zero and the total mass is very close to $2m_p$. To
evaluate the tunnelling rate we use the WKB method, however, since
the barrier is not much higher than the binding energy level,
using the WKB method might bring up certain errors. Therefore the
estimated width can only be valid to its order of magnitude.
Indeed, with present experimental accuracy, we can satisfy
ourselves with such numbers, but definitely the future experiments
can provide us with much more information and by them we will
modify our model and determine the concerned parameters to higher
accuracy.

The newly observed enhancement by BES and Belle may have various
interpretations, one of them is due to a resonance of $p\bar{p}$.
In this work we discuss this possibility in the linear
$\sigma-$model and the obtained values are quantitatively
consistent with the data. Our numerical results show, the total
width and position of the proposed bound state, no matter S-wave
or P-wave do not contradict the data, therefore both of them may
be possible states which can accommodate  the observed
enhancement.

The authors of \cite{Zou} suggested an alternative explanation,
i.e. the final state interaction results in the observed
enhancement. To decide which mechanism is right or dominant would
wait for the future experiments. We hope that studies on the new
resonance  can enrich our knowledge about the hadron physics and
the interactions at the hadron level. Our conclusion is that to
confirm  the observed enhancement, more precise measurements are
needed.

\vspace{1cm}

\noindent{Acknowledgment:}

We thank  K.T. Chao for his helpful comments and suggestions, we
also benefit from the  fruitful discussions with C.H. Chang. This
work is partially supported by the National Natural Science
Foundation of China.

\newpage
\begin{figure}[htb]
\begin{center}
\begin{tabular}{cc}
\scalebox{0.7}{\includegraphics{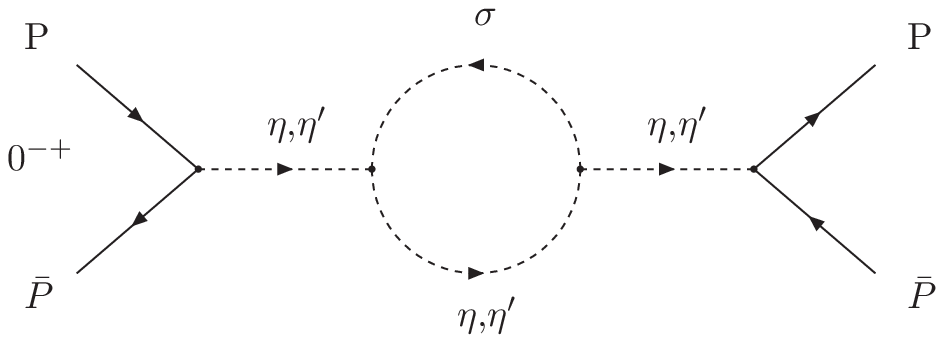}}&
\scalebox{0.7}{\includegraphics{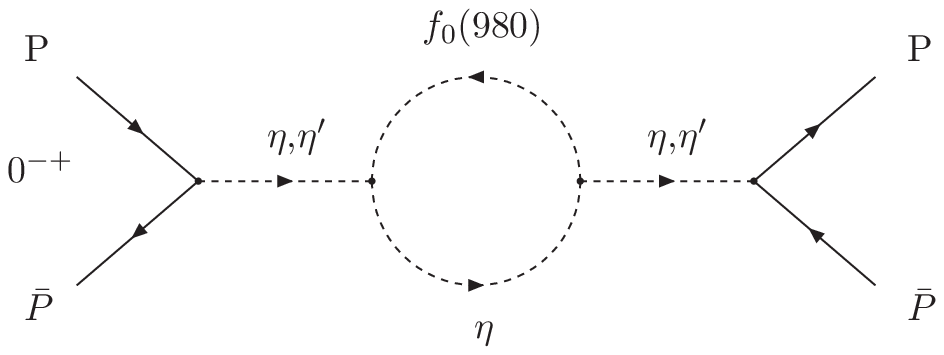}}\\\vspace{1cm}
(a)&(b)\\
\scalebox{0.7}{\includegraphics{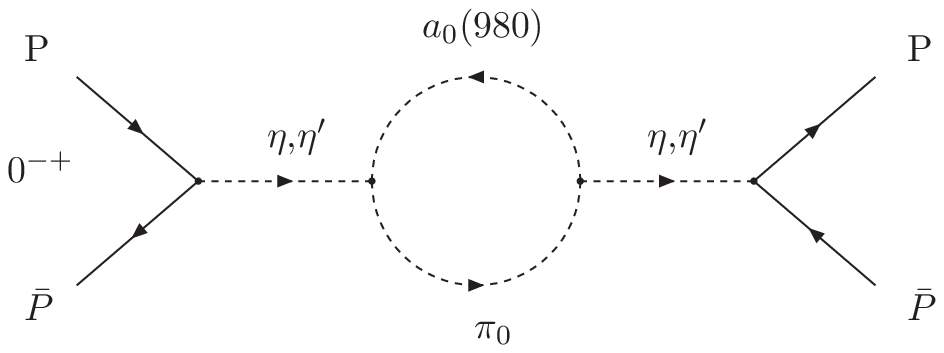}}&\scalebox{0.7}{\includegraphics{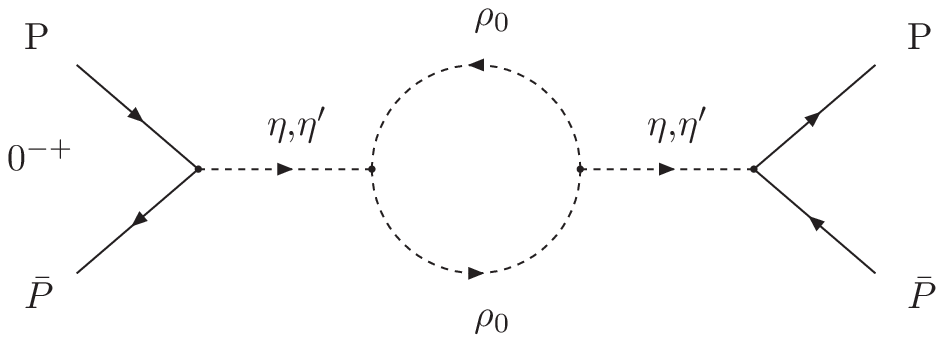}}
\\
(c)&(d)
\end{tabular}
\end{center}
\caption{}
\end{figure}

\begin{figure}[htb]
\begin{center}
\begin{tabular}{cc}
\scalebox{0.7}{\includegraphics{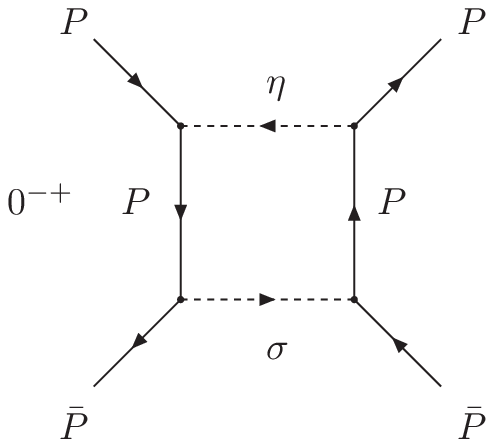}}\hspace{2cm}&\hspace{2cm}\scalebox{0.7}{\includegraphics{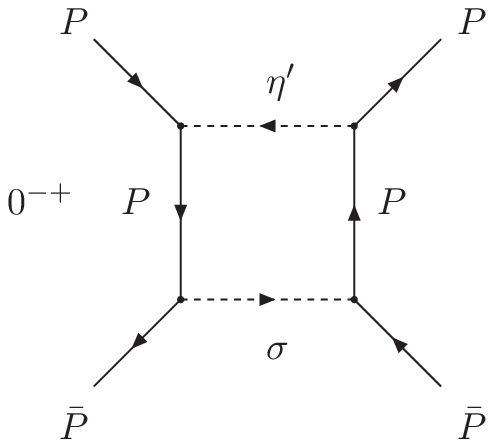}}\\
(a)\hspace{0.3cm}&\hspace{2.5cm}(b)\\
\end{tabular}
\end{center}
\caption{}
\end{figure}

\begin{figure}[htb]
\begin{center}
\scalebox{1.2}{\includegraphics{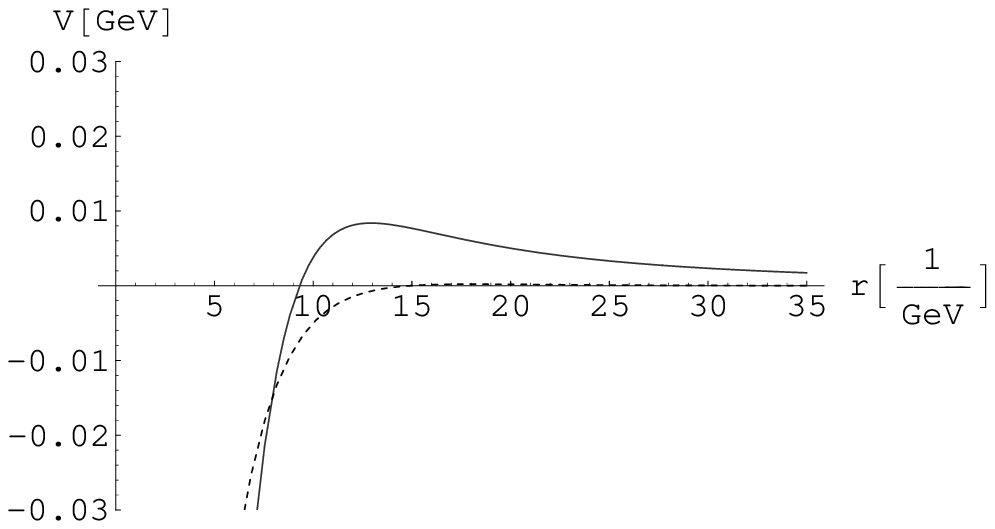}}
\end{center}
\caption{}
\end{figure}

\end{document}